\documentclass[showpacs,preprintnumbers]{revtex4}

\usepackage{mathrsfs}
\usepackage[latin1]{inputenc}
\usepackage[gen]{eurosym}
\usepackage{graphicx}
\usepackage{floatflt}
\usepackage{amsmath}
\usepackage{bbm}

\usepackage{color}

\usepackage{relsize}

\usepackage{fancybox}

\DeclareMathAlphabet{\mathpzc}{OT1}{pzc}{m}{it}

\usepackage{stmaryrd}

\usepackage{amssymb}

\usepackage{amsbsy}

\usepackage{amssymb}
\usepackage{cancel}

\usepackage{subfigure}

\usepackage[abs]{overpic}

\usepackage{amssymb}

\usepackage{epigraph}
\begin{document}

\title{Functional renormalization group study of the two-flavor linear
  sigma model in the presence of the axial anomaly}
\author{Mara Grahl$^{\text{(a)}}$ and Dirk H.\ Rischke$^{\text{(a,b)}}$}
\affiliation{$^{\text{(a)}}$Institute for Theoretical Physics, Goethe University,
Max-von-Laue-Str.\ 1, D--60438 Frankfurt am Main, Germany }
\affiliation{$^{\text{(b)}}$Frankfurt Institute for Advanced Studies, Goethe University,
Ruth-Moufang-Str.\ 1, D--60438 Frankfurt am Main, Germany }

\begin{abstract}
The $SU(2)_A \times U(2)_V$-symmetric chiral linear sigma model in
the presence of the axial anomaly is studied in the local-potential
approximation of the Functional Renormalization Group (FRG). The
renormalization group (RG) flow is investigated in a truncation which
reproduces recent results for the $U(2)_A \times U(2)_V$-symmetric
model in the limit of vanishing axial anomaly strength. We search for
the conjectured $O(4)$ fixed point in the presence of the $U(1)_A$ anomaly
and analyze its stability properties.        
\end{abstract}
\maketitle

\section{Introduction}

Pisarski and Wilczek \cite{Pisarski:1983ms} investigated the most 
general renormalizable Lagrangian which is invariant under the 
chiral $U(N_f)_L \times U(N_f)_R$ symmetry of quantum chromodynamics 
(QCD), where $N_f$ denotes the number of quark flavors. 
Choosing the $[\bar{N}_f,N_f]+[N_f,\bar{N}_f]$
representation of $SU(N_f)_L \times SU(N_f)_R$ \cite{Paterson:1980fc},
in Euclidean space this Lagrangian reads
\begin{gather}
 \mathscr{L}_\Phi = \frac{1}{2} {\rm Tr} (\partial_{\mu} \Phi^{\dagger} ) 
(\partial_{\mu} \Phi) +\frac{1}{2} m_{\Phi}^2 {\rm Tr} \Phi^{\dagger} \Phi 
+ \frac{\pi^2}{3} g_1 ({\rm Tr} \Phi^{\dagger} \Phi)^2 
+ \frac{\pi^2}{3} g_2 {\rm Tr} ( \Phi^{\dagger} \Phi)^2 \; , \label{lsp}
\end{gather}
where $\Phi$ is a complex-valued $N_f \times N_f$-matrix.
The anomalous breaking of the $U(1)_A$ symmetry contained in $U(N_f)_L
\times U(N_f)_R$ is due to instantons \cite{tHooft:1986} 
[see also Ref.\ \cite{Pawlowski:1996ch}] and
is commonly referred to as $U(1)_A$ anomaly. 
The authors of Ref.\ \cite{Pisarski:1983ms} conjectured that, 
for $N_f=2$, the chiral phase transition of QCD can be of
second order in the presence of the $U(1)_A$ anomaly. In this case, it
would fall into the $O(4)$ universality class. In the following, we
shall refer to this statement as \textit{O({\slshape 4}) conjecture}. \\
\noindent
The term commonly
introduced into Eq.\ (\ref{lsp}) in order to explicitly break the
$U(1)_A$ symmetry is
\begin{gather} \label{det}
\det{\Phi^{\dagger}} + \det{\Phi}\;.
\end{gather}
In Appendix \ref{AppA} we show that, for $N_f=2$, the most general
form of the anomaly including terms up to naive scaling dimension four is
\cite{Butti:2003nu}
\begin{gather}
{\cal L}_{A} =
c \left(\det{\Phi^{\dagger}} + \det{\Phi} \right) + y
\left(\det{\Phi^{\dagger}} + \det{\Phi} \right) {\rm Tr}
\Phi^{\dagger} \Phi  + z \left[ \left( \det{\Phi^{\dagger}} \right)^2 
+ \left( \det{\Phi} \right)^2 \right] \; . \label{ano}
\end{gather}
These terms must be added to Eq.\ (\ref{lsp}),
\begin{gather} \label{ano2}
{\cal L} = {\cal L}_\Phi + {\cal L}_{A}\;,
\end{gather}
if one wants to study the impact of the $U(1)_A$ anomaly on the chiral
phase transition.  For $N_f=2$ and including terms up to naive scaling
dimension four, the Lagrangian (\ref{ano2})
is the most general Lagrangian invariant under $SU(2)_A \times U(2)_V$
and respecting parity symmetry. 
We note that the terms $\sim y, z$ are always
induced by the RG flow if $c \neq 0$. Therefore, in the following
we shall use the notion ``in the presence of the
anomaly'', whenever $c \neq 0$. Note also that 
\begin{gather}
\label{dep}
 \left(\det{\Phi^{\dagger}} + \det{\Phi} \right)^2=-{\rm Tr} (
 \Phi^{\dagger} \Phi)^2
+({\rm Tr} \Phi^{\dagger} \Phi)^2+\left[ \left( \det{\Phi^{\dagger}}
  \right)^2 + \left( \det{\Phi} \right)^2 \right] \; ,
\end{gather}
so that the square of the term (\ref{det}) is not linearly independent
from the other invariants contained in Eq.\ (\ref{ano}). Finally note that
\begin{gather}
 i \left(\det{\Phi^{\dagger}} - \det{\Phi} \right)
\end{gather}
is not invariant under $CP$ transformations \cite{Pisarski:1983ms}. \\
\noindent
In this work, we consider the case $N_f=2$. Denoting
\begin{gather}
\Phi = \left(\sigma + i \eta \right) t_0 + \vec{t} \cdot \left(\vec{a}
  + i \vec{\pi} \right)\;,
\end{gather}
with $t_0= \frac{1}{\sqrt{2}} \begin{pmatrix} 1&0\\ 0&1
\end{pmatrix}$, 
$t_1= \frac{1}{\sqrt{2}} \begin{pmatrix} 0&1\\ 1&0 \end{pmatrix}$, 
$t_2= \frac{1}{\sqrt{2}} \begin{pmatrix} 0&-i\\ i&0 \end{pmatrix}$, 
$t_3= \frac{1}{\sqrt{2}} \begin{pmatrix} 1&0\\ 0&-1 \end{pmatrix}$, we
rewrite the Lagrangian (\ref{ano2}) into the form \cite{Fukushima:2010ji}
\begin{gather}
 \mathscr{L}=\frac{1}{2} \left(\partial_{\mu} \sigma \partial_{\mu} \sigma 
+\partial_{\mu} \vec{\pi} \cdot \partial_{\mu} \vec{\pi} 
+\partial_{\mu} \eta \partial_{\mu} \eta 
+\partial_{\mu} \vec{a} \cdot \partial_{\mu} \vec{a} \right) + U \; , 
\label{lsl} \\
U= \frac{1}{2} \mu^2 \left(\sigma^2 + \vec{\pi}^2 + \eta^2 + \vec{a}^2 \right)
 + \frac{\lambda_1}{4!} \left(\sigma^2 + \vec{\pi}^2 + \eta^2 +
   \vec{a}^2 \right)^2     \nonumber \\       
+ \lambda_2 \left[ \left(\sigma^2 +\vec{\pi}^2 \right) \left(\eta^2 +
    \vec{a}^2 \right) 
- \left(\sigma \eta - \vec{\pi} \cdot \vec{a} \right)^2  \right] \nonumber \\
 +c \left(\sigma^2 -\eta^2 + \vec{\pi}^2-\vec{a}^2 \right)
+ y \left(\sigma^2+ \vec{\pi}^2 + \eta^2 + \vec{a}^2 \right) 
\left(\sigma^2-\eta^2 + \vec{\pi}^2 -\vec{a}^2 \right)   \nonumber \\
 +z \frac{1}{2} \left(\eta^2+\vec{a}^2-\sigma^2-\vec{\pi}^2-2\vec{a}
   \cdot \vec{\pi}+2\eta\sigma \right)
\left(\eta^2+\vec{a}^2-\sigma^2-\vec{\pi}^2+2\vec{a} \cdot
  \vec{\pi}-2\eta\sigma \right) 
\; ,  \label{ls}
\end{gather}
where $\lambda_1 \equiv 4! \frac{\pi^2}{3} \left(g_1 + \frac{1}{2} g_2
\right)$, $\lambda_2 \equiv 2 \frac{\pi^2}{3} g_2$, $\mu^2 \equiv
m_{\Phi}^2$.
For $c=0$, $y=0$, and $z=0$ Eq.\ (\ref{lsl}) reduces to the 
$U(2)_L \times U(2)_R$-symmetric Lagrangian (\ref{lsp}).\\
\noindent
The RG flow for the Lagrangian (\ref{lsp}) was analyzed for different
values of $N_f$. The results from the $\epsilon$-expansion 
\cite{Butti:2003nu,Pisarski:1980ix} prove that for $N_f=2$ the
$O(8)$-symmetric infrared (IR) fixed point is unstable, which is confirmed 
from FRG studies \cite{Fukushima:2010ji} as well as from lattice
calculations \cite{Espriu:1997vc}. The absence of a IR stable 
fixed point is a sufficient criterion for the phase transition to be
of first order. In the presence of the anomaly ($c \ne 0$), however, 
to our knowledge the RG flow for the Lagrangian (\ref{lsl}) has not
yet been calculated explicitly, neither in the $\epsilon$-expansion
nor in the FRG framework. The FRG study presented in Refs.\ 
\cite{Berges:1997eu,Berges:1998sd} neglects the fields $\eta$ and
$\vec{a}$ from the beginning. Other RG results in the presence
of the anomaly can be found in the literature only for cases where the
anomaly term acts as a coupling of order higher than two [see for example
Refs.\ \cite{Wirstam:2002be,BJnf3,Jiang:2012wm,Patkos:2012ex}]. Also, studying how $c$ approaches $\infty$
has not yet been investigated explicitly on the level of RG flow
equations. In this work we want to fill these gaps by appropriately
extending the study presented in Ref.\ \cite{Fukushima:2010ji}. \\
\noindent
For the remainder of this introductory section, we would like to make
a couple of remarks. The first one concerns the 
\emph{universality hypothesis}. The RG approach towards critical phenomena  
defines (universality) classes of microscopically very different
models, which lie in the basin of attraction of a certain
IR stable fixed point and hence share the
same critical exponents. Each universality class can be uniquely
defined by the universal eigenvalues of the stability matrix associated with the 
IR stable fixed point [we neglect cases where the
critical exponents depend on the couplings, as in Baxter's famous
two-dimensional eight-vertex model \cite{Baxter:1989}]. Also the fixed-point
potential is characteristic for a given universality class and determines 
the symmetry of the fixed point. \\
\noindent
Consider a certain symmetry group $G_0$ and
the most general $G_0$-invariant Landau-Wilson potential
for a certain representation $\Gamma(G_0)$.
For the sake of simplicity let us assume that 
the Landau-Wilson potential has a single IR stable fixed
point $FP_0$ in coupling space, and thus falls into the universality
class of $FP_0$. \\
\noindent
Let us now add another coupling term 
to the potential, which breaks $G_0$ to a subgroup $G$ of $G_0$.
One obtains a different model which is only invariant 
under this subgroup $G \subset G_0$.
The presence of an additional coupling term could induce
another IR stable fixed point $FP$ associated with $G$.
The existence of the fixed point $FP_0$ need not be compromised, but it
does not need to be IR stable anymore. 
However, if $FP_0$ remains IR
stable, and if $FP$ does not exist, and if no separatrix exists in the RG flow, 
the new $G$-invariant model necessarily falls into the same
universality class as the previous $G_0$-invariant model.\\
\noindent 
In the literature there exist different versions of the
universality hypothesis with slightly varying scope and content 
\cite{PhysRevLett.24.1479,Bruce:1980,Baxter:1989}. 
We state the universality hypothesis as follows: two Lagrangians 
for two different order parameters lie in the same universality class, if
(i) the spatial dimension is the same for
both systems, (ii) the order parameters have the same number of
components, (iii) the symmetries of the Lagrangians are isomorphic, 
and (iv) there are no long-range interactions in both Lagrangians. [Usually, long-range
interactions yield mean-field values for the critical exponents and
one does not have nontrivial universal behavior. In the presence of
``middle-range'' interactions critical exponents can be different for 
two Lagrangians fulfilling the criteria (i) -- (iii).] Whereas conditions
(i), (ii), and (iv) are necessary conditions, criterion
(iii) is sufficient but not
necessary since, according to the above discussion, the fixed point
for the full symmetry group $G_0$ can remain IR
stable even in the presence of terms which break the symmetry to $G \subset
G_0$, and therefore the two Lagrangians are both in the 
universality class associated with $FP_0$. 
It is an open question how to turn (iii) into a
necessary condition, and if further conditions are necessary in order to
exclude exceptions \cite{Bruce:1980}. \\
\noindent
On the other hand, there exists a plethora of (more or less reliable)
criteria which can serve to rule out the existence of an IR stable fixed
point [see, for example, Refs.\
\cite{Toledano:1985,ToledanoLandau,Sen:2010,Friman1} and references therein]. The
best-known ones were already given by Landau and Lifshitz \cite{LL,ToledanoLandau}, namely
the case where the 
representation $\Gamma(G)$ of the group associated with the
$G$-invariant Landau-Wilson potential is not irreducible (such that
there is a linear invariant), or the case where the third power of the
representation, $\Gamma^3(G)$, 
contains the trivial representation (such that there is a third-order invariant
which drives the transition first order). 

\noindent
Furthermore, unless one is interested in multicritical
behavior [for a related work, see Ref.\ \cite{Eichhorn:2013zza}],
it is commonly assumed \cite{Krinsky} that in case of two 
quadratic invariants, one can restrict the discussion of critical
behavior near second-order phase transitions to simpler models, one
for each of the competing order parameters (the quadratic invariants).
This is because in general the couplings associated to the two quadratic invariants
vanish at different critical temperatures, each corresponding to a different phase
transition. One may naively think that it
should therefore be possible to ignore one of the order parameters, when
discussing the phase transition for the other one. However, the second
invariant introduces another relevant direction in coupling space
which may render an IR stable fixed point corresponding to
a second-order phase transition associated with one of the order
parameters unstable. \\
\noindent
Our second remark concerns the role of baryon number conservation in
the chiral phase transition. 
The $U(1)_A$ anomaly 
explicitly breaks the $U(1)_A$ symmetry contained in $G \equiv U(N_f)_V
\times U(N_f)_A \simeq U(1)_V \times U(1)_A \times 
\left[ SU(N_f)/Z(N_f) \right]_L \times \left[ SU(N_f)/Z(N_f)
\right]_R$ down to $Z(N_f)_A$, where $\simeq$ symbolizes group
isomorphy. The group $U(1)_V$ is associated with
baryon number conservation and should not be broken (spontaneously) 
during the phase transition. Thus one usually argues that one can
neglect it when studying the chiral phase transition, leaving 
$\left[ SU(N_f)_L \times SU(N_f)_R \right] / Z(N_f)_V 
\rightarrow SU(N_f)_V / Z(N_f)_V $
for the symmetry breaking pattern relevant for the chiral phase
transition in the presence of the anomaly \cite{Butti:2003nu}. 
The spontaneous breaking of a discrete symmetry does not yield 
Goldstone modes, such that it is sufficient to consider the breaking of the
continuous group $G' \equiv SU(N_f)_L \times SU(N_f)_R$ in the chiral phase 
transition in the presence of the anomaly. We nevertheless
consider an effective theory for the order parameter
invariant under $U(1)_V \times G'$ in the search for the IR
fixed point associated to spontaneous breaking of $SU(N_f)_L \times
SU(N_f)_R$. \\
\noindent
Our final remark concerns the \textit{O({\slshape 4}) conjecture}.
Aside from criterion (iv) which we do not discuss here, 
we conclude that if the chiral phase transition of two-flavor QCD in
the presence of the anomaly is of second order, then the Lagrangian 
(\ref{lsl}) should fall into the same universality class as QCD. 
The Lagrangian (\ref{ano2}) however has eight degrees of freedom, whereas the 
$O(4)$ model has only four, which at first glance would mean that 
criterion (ii) of
the universality hypothesis is not fulfilled. It is therefore a priori
not clear that the IR stable fixed point for the Lagrangian 
(\ref{ano2}) is an $O(4)$ fixed point. It might as well correspond
to another universality class, characterized by $SU(2)_A\times U(2)_V$
critical exponents.
To justify the \textit{O({\slshape 4}) conjecture}, first note that the choice
of the representation depends on the physical degrees of 
freedom one intends to study.
In the presence of the anomaly, one can make use of the isomorphism
\begin{gather}
SU(2) \times SU(2)/Z(2) \simeq SO(4)\;,
\end{gather}
which means that $SU(2) \times SU(2)$ is locally isomorphic to $O(4)$. 
Accordingly, $SU(2) \times SU(2)$ has an $O(4)$ representation. 
For $N_f =2$, the representation of the Lagrangian (\ref{lsl}), or 
(\ref{ano2}), respectively, is reducible. 
It consists of the sum of two equivalent  $O(4)$ representations
\cite{Paterson:1980fc,Jungnickel:1995fp,Butti:2003nu}, 
$\Phi_1 = \sigma t_0 + i \vec{t} \cdot \vec{\pi}$ and 
$\Phi_2 =  i \eta t_0 + \vec{t} \cdot \vec{a} $, which are both
irreducible, but not faithful, representations of $SU(2) \times SU(2)$. 
Therefore, the symmetry of QCD allows for an $O(4)$ representation, 
if only the sigma and pion are light particles.
At mean-field level this can be confirmed. 
The analysis in Refs.\ \cite{'tHooft:1986nc,Yagi:2005yb}
shows that if we identify $\vec{\pi}$ with the Goldstone 
modes (the pions), the fields $\eta$ and $\vec{a}$ are massive at the 
critical point, whereas the field $\sigma$ is as light as the pions (and can be 
interpreted as the chiral partner of the pion). Since at the critical 
point only the modes with smallest mass are relevant (i.e., which 
count as components of the order parameter), we conclude that, if the
mean-field approximation were justified, the IR fixed point would indeed be the stable 
Wilson-Fisher fixed point of the $O(4)$ model.    \\
\noindent
Of course, the mean-field approximation neglects quantum fluctuations
(such as instantons), which might change the universality class or might lead to
the instability of the fixed point. For this reason we study the 
FRG flow for the Lagrangian
(\ref{lsl}) in this paper. One could argue that for very large 
anomaly strength, $c \rightarrow -\infty$, $\eta$- and $\vec{a}$-loop
diagrams should be suppressed according to the Appelquist-Carazzone 
decoupling theorem \cite{AppelquistCarazzone} due to the very high tree-level mass 
for the corresponding fields. Since the $\epsilon$-expansion deals 
only with loop diagrams, one can indeed expect to find the 
$O(4)$ fixed point 
\footnote{Note that in the limit $c \rightarrow -\infty$ the $O(4)$ 
fixed point corresponds to the above mentioned $O(4)$ representation 
$\Phi_1 = \sigma t_0 + i \vec{t} \cdot \vec{\pi}$. The limit 
$c \rightarrow \infty$ would in turn correspond to the equivalent
$O(4)$ representation $\Phi_2 =  i \eta t_0 + \vec{t} \cdot \vec{a} $ 
with $\sigma$ and $\vec{\pi}$ simply exchanging roles with $\eta$ and
$\vec{a}$.}. 
However, this argument says nothing about (a) the stability of the 
$O(4)$ fixed point and (b) the cases of small and 
intermediate anomaly strength. \\
\noindent
Let us note that in consistency with Refs.\ 
\cite{Pisarski:1983ms,Fukushima:2010ji} we work with the 
dimensionally reduced theory, which is justified because the diverging
correlation length at a second-order phase transition leads to dimensional 
reduction \cite{DimRed2ndorder}. Again in agreement with the 
aforementioned references, we restrict ourselves to the 
case $T \gtrsim T_c$ where the system is driven towards the critical 
point from the side of the restored phase. This allows to assume 
vanishing vacuum expectation values for all fields.  \\
\noindent
As a final remark, we want to point the reader to several FRG studies related
to our work in a larger context, the list of which is, however, incomplete.
Here we want to mention the recent work on applying the FRG method to
QCD \cite{Braun:2006jd,Braun:2009gm,Pawlowski:2010ht,Braun:2011pp,Li:2012fe},
a strategy how to combine
first-principle QCD flows with effective models \cite{Haas:2013qwp},
and some investigations of effective models for QCD 
\cite{Schaefer:2006sr,Herbst:2013ail}. \\
\noindent  
This paper is organized as follows. At the beginning of Sec.\
\ref{sec2} we explain the method we use. In Secs.\ \ref{sing} and
\ref{secz}, respectively, we consider two equivalent parameterizations
of the potential. This not only serves as a check of our results, but
also illustrates our general remarks given in Appendix \ref{o8} on how
to obtain the correct flow equations when working with a
parameterization in terms of the original field components instead of
invariants. We derive the flow equations and analyze the stability
properties of the fixed points. In Sec.\ \ref{pan} we explain why the 
$O(4)$ fixed point becomes stable in the special case of infinite 
anomaly strength. Our arguments are supported by a discussion of the 
analogous situation in a simpler model in Sec.\ \ref{coupvec}.
Section \ref{conclusions} concludes this work with a summary of our results.

\section{Linear sigma model from FRG}
\label{sec2}

In this section we investigate the FRG flow of the Lagrangian
(\ref{lsl}) in different parameterizations proceeding in analogy 
to Ref.\ \cite{Fukushima:2010ji}. We reproduce 
the result of Ref.\ \cite{Fukushima:2010ji} in the limit 
$c,y,z \rightarrow 0$. \\
\noindent
We use the FRG equation in the local-potential approximation:
\begin{gather}
 \frac{\partial U_k}{\partial k} = K_d k^{d+1} \sum_{i}
 \frac{1}{E_i^2} \; , \; \; E_i^2 \equiv k^2 + M_i^2 \; , 
\; \; K_d \equiv \frac{2\pi^{d/2}}{d \ \Gamma(d/2) (2\pi)^d} \; , \label{fleq}
\end{gather}
\noindent
where $\mathscr{L}_{k} = \frac{1}{2} Tr (\partial_{\mu}
\Phi^{\dagger}) (\partial_{\mu} \Phi)+U_k$,  
with $\mathscr{L}_{k=\Lambda}=\mathscr{L}$ being the bare Lagrangian,
and $d$ is the spatial dimension. 
In the following, all our numerical results are for $d=3$.
$M_i^2$ denote the eigenvalues of the mass matrix
\begin{gather}
 M_{ij} \equiv \frac{\partial^2 U_k}{\partial \phi_i \partial \phi_j}
 \; , \; \; i,j=1,\ldots,8 \; , \label{mm}
\end{gather}
\noindent
where the fields $\phi_i$ are given by $\sigma,\vec{\pi},\eta,$ 
and $\vec{a}$. \\
\noindent
With the invariants 
\begin{gather}
\label{abb}
\varphi \equiv \sigma^2+\vec{\pi}^2+\eta^2+\vec{a}^2 \; , \ 
\ \xi = (\sigma^2+\vec{\pi}^2)(\eta^2+\vec{a}^2)-(\sigma \eta - 
\vec{\pi} \cdot \vec{a})^2 \; , \; \; 
\alpha \equiv \sigma^2-\eta^2+\vec{\pi}^2-\vec{a}^2 \; ,
\end{gather}
\noindent
and the abbreviation
\begin{gather*}
\beta \equiv \alpha^2 -\frac{\varphi^2}{2} + 2\xi = \frac{1}{2} \left(\eta^2+\vec{a}^2-\sigma^2
-\vec{\pi}^2-2\vec{a} \cdot \vec{\pi}+2\eta\sigma \right)
\left(\eta^2+\vec{a}^2-\sigma^2-\vec{\pi}^2+2\vec{a} \cdot 
\vec{\pi}-2\eta\sigma \right) \; ,
\end{gather*}
\noindent
the bare potential (\ref{ls}) reads 
\begin{gather}
\label{pold}
 U (\varphi,\xi,\alpha) =  \frac{1}{2} \mu^2 \varphi + \frac{1}{4!}
 \lambda_1 \varphi^2 + \lambda_2 \xi + c \alpha + y \alpha \varphi + z \beta\; .
\end{gather}
\noindent
Using relation (\ref{dep}) and a different notation,
\begin{gather}
 \varphi_1 = \sigma^2+\vec{\pi}^2 \; , \; \; 
\varphi_2 = \eta^2+\vec{a}^2 \; , \; \; 
\gamma = (\sigma \eta - \vec{\pi} \cdot \vec{a})^2 \; ,
\end{gather}
\noindent
we obtain
\begin{gather}
\label{pinv}
U(\varphi_1, \varphi_2, \gamma) =  m_1^2 \varphi_1 + m_2^2 \varphi_2 +
l_1 \varphi_1^2 + l_2 \varphi_2^2 + l_{12} \varphi_1 \varphi_2 +l_3 \gamma \; ,
\end{gather}
\noindent
where we introduced new couplings,
\begin{gather}
m_1^2 = \frac{1}{2} \mu^2 + c \; , \; \; m_2^2 = \frac{1}{2} \mu^2 -c \ , \\
 l_1 = y + \frac{\lambda_1}{4!} + \frac{z}{2} \ , \ \ 
l_2 = -y + \frac{\lambda_1}{4!} + \frac{z}{2} \ , \ \ 
l_{12} = \frac{\lambda_1}{12}+\lambda_2 -z  \ , \ \ 
l_3 = -(\lambda_2 + 2z) \ .
\end{gather}
\noindent
Note that the number of linearly independent invariants is the same in expressions
(\ref{pold}) and (\ref{pinv}), respectively. 
\noindent
When calculating the mass eigenvalues $M_i$, we have to simplify the 
computation by setting the values of several fields to zero after 
having performed the second derivatives in Eq.\ (\ref{mm}). 
Keeping all fields nonzero, we obtain complicated expressions 
for the eigenvalues because an $8 \times 8$ matrix has to be
diagonalized. 
One can circumvent the diagonalization using the relation
\begin{gather}
 \sum_{i} \frac{1}{k^2 + M_i^2} = Tr \mathcal{M}^{-1} \ , \ \ 
\mathcal{M}_{ij} \equiv M_{ij} +k^2 \delta_{ij} \ . 
\end{gather}
\noindent
However, it still would take a symbolic computation program a long
time to expand the r.h.s.\ of the FRG equation (\ref{fleq}) in powers 
of the fields.
Fortunately, the $\epsilon$-expansion results from Ref.\  
\cite{Pisarski:1983ms} can be reproduced by keeping nonzero
values only for $\sigma$ and one of the components of $\vec{a}$, 
say $a_1$ \cite{Fukushima:2010ji}. Note that this is not possible 
if we choose another field than a component of $\vec{a}$, since 
then $\xi = 0$ and we do not obtain a flow equation for $\lambda_2$. 
We further comment on the validity of this procedure from a more 
general perspective in Appendix \ref{o8}.

\subsection{Parameterization in terms of invariants}
\label{sing}

In this section we use the parameterization (\ref{pinv}) for the
potential. It is nontrivial to rewrite all fields
$\phi_i$ in terms of the above invariants. 
Since we have three invariants, the rewriting can be performed 
unambiguously only if we keep at least three fields nonzero.
Keeping $\eta$, $\sigma$, and $a_1$ nonzero, we obtain 
the unambiguous mapping
\begin{gather}
 \sigma = \sqrt{\varphi_1} \ , \ \ a_1 = \sqrt{\frac{\varphi_1
     \varphi_2 - \gamma}{\varphi_1}} \ , \ \ 
\eta = \sqrt{\frac{\gamma}{\varphi_1}} \ .
\end{gather}
\noindent
We also repeated our analysis using $\pi_1$ instead of $\eta$ and 
found identical results. \\
\noindent
We express the mass eigenvalues $M_i$ in terms of 
$\varphi_1$, $\varphi_2$, and $\gamma$ and
expand the r.h.s.\ of Eq.\ (\ref{fleq}) in powers of these invariants.
Then, inserting  
Eq.\ (\ref{pinv}) on the l.h.s., we read off flow equations 
for the couplings by comparing coefficients.
In order to calculate critical exponents we rescale quantities 
to obtain flow equations for dimensionless parameters. With 
\begin{gather}
 m_{i,k}^2 = k^2 \bar{m}_{i,k}^2 \ , \ \  l_{i,k} =  k^{4-d} \bar{l}_{i,k} \ ,
\end{gather}
\noindent
we obtain
\begin{gather}
k  \frac{\partial \bar{m}_1^2 }{\partial k}
=  -2 \bar{m}_1^2-\frac{1}{3 \pi ^2}
\left(\frac{12 \bar{l}_1}{\bar{\epsilon }_1^2}+\frac{\bar{l}_3
+4 \bar{l}_{12}}{\bar{\epsilon }_2^2}\right) \ , \\ \label{flf}
 k  \frac{\partial \bar{m}_2^2 }{\partial k}
= -2 \bar{m}_2^2-\frac{1}{3 \pi ^2}\left(\frac{\bar{l}_3
+4 \bar{l}_{12}}{\bar{\epsilon }_1^2}
+\frac{12 \bar{l}_2}{\bar{\epsilon }_2^2}\right) \ , \\
  k  \frac{\partial \bar{l}_{12} }{\partial k} 
=    -\bar{l}_{12}+\frac{2 \left[  4( \bar{l}_1  \bar{\epsilon }_2^3 
+  \bar{l}_2  \bar{\epsilon }_1^3) \left(\bar{l}_3
+6 \bar{l}_{12}\right)+ \left(\bar{l}_3^2+4 \bar{l}_{12}^2\right)
   \left(\bar{\epsilon}_1+\bar{\epsilon}_2\right) \bar{\epsilon }_1 
\bar{\epsilon }_2\right]}{3 \pi ^2
   \bar{\epsilon }_1^3 \bar{\epsilon }_2^3} \ , \\
 k  \frac{\partial \bar{l}_{1} }{\partial k} 
=  -\bar{l}_1+\frac{2}{3 \pi ^2}\left(\frac{48
    \bar{l}_1^2}{\bar{\epsilon }_1^3}
+\frac{\bar{l}_3^2+2 \bar{l}_3 \bar{l}_{12}
+4 \bar{l}_{12}^2}{\bar{\epsilon }_2^3}\right) \ , \\
 k  \frac{\partial \bar{l}_{2} }{\partial k} 
=   -\bar{l}_2+\frac{2}{3 \pi ^2}\left(   
\frac{48 \bar{l}_2^2}{\bar{\epsilon }_2^3} 
+ \frac{\bar{l}_3^2+2 \bar{l}_3 \bar{l}_{12}
+4 \bar{l}_{12}^2}{\bar{\epsilon }_1^3} \right) \ , \\
  k  \frac{\partial \bar{l}_{3} }{\partial k} =
 -\bar{l}_3+\frac{4 \bar{l}_3 \left[4 \bar{l}_1 \bar{\epsilon }_2^3
+ 4 \bar{l}_2 \bar{\epsilon }_1^3+\left(3 \bar{l}_3+4 \bar{l}_{12} \right)
   \left(\bar{\epsilon}_1+\bar{\epsilon}_2\right)\bar{\epsilon }_1 
\bar{\epsilon }_2\right]}{3 \pi ^2 \bar{\epsilon }_1^3 
\bar{\epsilon }_2^3} \ , \label{fll}
\end{gather}
\noindent
where we omitted the index $k$ and used the abbreviation
\begin{gather}
\label{abbr}
 \bar{\epsilon}_i = 1+2  \bar{m}_i^2 \ .
\end{gather}
\noindent
In order to find the fixed points we have to set the left-hand sides to zero 
and solve the resulting system of equations. Since the equations 
are nonlinear, this has to be done numerically, using starting values 
for which a standard root-finding algorithm converges towards a
solution. 
We applied an algorithm with randomized starting values in a 
reasonably large domain of parameter space, each of the starting
values lying in the interval $[-10^4,10^4]$. We found
the nontrivial solutions given in Table \ref{tabfp} where they
are listed together with the corresponding eigenvalues of the 
stability matrix. Using $10^6$ different starting values,
we checked that these are the only solutions for given starting values
in the above domain of parameter space.\\
\noindent
The method how to calculate the eigenvalues of the stability matrix 
is described in Appendix \ref{crex}. From comparison with the
corresponding eigenvalues for 
the $O(N)$ models in the same approximation scheme 
(local-potential approximation, fourth-order truncation in the
fields), see Appendix \ref{o8}, we can unambiguously identify those
fixed points in Table \ref{tabfp} with $O(N)$ 
critical exponents. Let us start our discussion with 
the fixed points $FP_6$ and $FP_7$. 
From the vanishing of the couplings in
the upper part of Table \ref{tabfp} we see that for each of
these fixed points the
fixed-point potential is that of an $O(4)$ model. Fixed point
$FP_6$ is that for the
$O(4)$ representation $\Phi_1 = \sigma t_0 + i \vec{t} \cdot
\vec{\pi}$, while $FP_7$ that for 
$\Phi_2 =  i \eta t_0 + \vec{t} \cdot \vec{a}$. 
From the eigenvalues of the
stability matrix in the lower part of Table \ref{tabfp}
one observes that both fixed points 
have more than one negative eigenvalue, which means
that they are unstable. Comparison of the second
and third eigenvalue with the last two columns in Table \ref{tab1} of Appendix
\ref{o8} also tells us that they have one relevant $O(4)$ scaling direction.\\
\noindent
For fixed point $FP_5$, the two masses $\bar{m}_i^2$ and the 
two coupling constants $\bar{l}_i$ are identical, while 
$\bar{l}_{12}=0$. This means that the fixed-point potential 
is that of two independent, identical $O(4)$ models. 
From the lower part of Table \ref{tabfp}
we see that this fixed point is a multicritical fixed point with 
two relevant $O(4)$ scaling directions. The third negative eigenvalue
of the stability matrix renders this an unstable fixed point.
Fixed point $FP_8$ is another 
unstable multicritical fixed point with a single $O(4)$ scaling
direction. \\
\noindent
From the vanishing of $\bar{l}_3$ and the fact that
$\bar{l}_{1}= \bar{l}_2=\bar{l}_{12}/2$, the fixed-point potential
for $FP_9$ is that of an $O(8)$ model. The stability matrix
indicates that this fixed point is unstable. Comparison of
the eigenvalues of the stability matrix with Table
\ref{tab1} shows that it has one relevant $O(8)$ scaling direction.
Since all eigenvalues of the stability matrix are negative,
fixed points $FP_{10}$ and $FP_{11}$ are ultraviolet (UV) stable 
fixed points.
Fixed points $FP_1$ and $FP_2$ are unstable fixed points, none of them
belonging to one of the $O(N)$ universality classes. \\
\noindent
Finally, $FP_3$ and $FP_4$ are IR
stable fixed points. While for the
other fixed points the (rescaled) eigenvalues of the (squared) mass matrix
are always positive semi-definite, for $FP_3$ and $FP_4$
we find one negative eigenvalue in all minima of the (rescaled)
fixed-point potential $\bar{U}(\bar{\sigma},\bar{\eta},\bar{a}_1)$, 
which corresponds to an unphysical
situation. However, this could be an artefact of our
fourth-order truncation of the potential (\ref{pinv}), and
the masses could be real-valued in higher order \cite{Marainprep}.
In that case, these IR stable fixed points are in the
$SU(2) \times U(2)$ universality class. Nevertheless,
at our truncation order we have to reject them.

\begin{table}
\caption{\label{tabfp}Fixed points in the presence of nonzero
anomaly strength, in $d=3$ dimension, in the FRG analysis in
the local-potential approximation, with couplings up to quartic order. 
The bar denotes rescaled quantities. }
\begin{ruledtabular}
\begin{tabular}{l|l l l l l l  }
 FP & $\bar{m}_1^2$ & $\bar{m}_2^2$ & $\bar{l}_1$ & $\bar{l}_2$ &
 $\bar{l}_{12}$ & $\bar{l}_3$   \\ \hline
 $FP_1$ & -3.80278 & -0.197224 & -355.58 & 0.273944 & 29.6088 & 0 \\
 $FP_2$ &  -0.197224 & -3.80278 & 0.273944 & -355.58 & 29.6088 & 0 \\
 $FP_3$ & -1.34694 & -0.333929 & -17.0334 & 0.128417 & 5.64724 & -5.93079  \\
 $FP_4$ &  -0.333929 & -1.34694 & 0.128417 & -17.0334 & 5.64724 & -5.93079 \\
 $FP_5$ & -0.0555556 & -0.0555556 & 0.216617 & 0.216617 & 0 & 0  \\
 $FP_6$ &   -0.0555556 & 0 & 0.216617 & 0 & 0 & 0 \\
 $FP_7$ & 0 & -0.0555556 & 0 & 0.216617 & 0 & 0  \\
 $FP_8$ & -0.0555556 & -0.0555556 & 0.108308 & 0.108308 & 0.216617 & 0.433234 \\
 $FP_9$ & -0.0675676 & -0.0675676 & 0.149643 & 0.149643 & 0.299286 & 0  \\
 $FP_{10}$ & 0.609013 & -1.18037 & -34.5716 & 7.9826 & -9.98504 & 129.304 \\
 $FP_{11}$ &  -1.18037 & 0.609013 & 7.9826 & -34.5716 & -9.98504 & 129.304  \\
\end{tabular}
\end{ruledtabular}
\begin{ruledtabular}
\begin{tabular}{l|c  }
 FP & stability-matrix eigenvalues   \\ \hline
 $FP_1$ &  \{9.64793,-5.66667,-0.585909+4.07239 i,-0.585909-4.07239
   i,3.83241,-1.30852\} \\
 $FP_2$  & \{9.64793,-5.66667,-0.585909+4.07239 i,-0.585909-4.07239
   i,3.83241,-1.30852\} \\
 $FP_3$ & \{29.6235,14.0524 +4.23653 i,14.0524 -4.23653 i,0.917927
 +9.64911 i,0.917927 -9.64911 i,-1.17232\} \\
 $FP_4$  & \{29.6235,14.0524 +4.23653 i,14.0524 -4.23653 i,0.917927
 +9.64911 i,0.917927 -9.64911 i,-1.17232\} \\
 $FP_5$  & \{-1.77069,-1.77069,1.27069,1.27069,-0.666667,0 \} \\
 $FP_6$  & \{-2.,-1.77069,1.27069,-1.,-0.833333,-0.5\} \\
 $FP_7$  & \{-2.,-1.77069,1.27069,-1.,-0.833333,-0.5\} \\
 $FP_8$  &  \{-2.,-1.77069,1.27069,-0.666667,0,0\} \\
 $FP_9$  & \{-1.98804,-1.71971,1.34471,0.613041,-0.25,-0.25\} \\
 $FP_{10}$  & \{-28.9145,-16.865,-10.9156,-3.11604+5.87462
 i,-3.11604-5.87462 i,-1.28288\} \\
 $FP_{11}$ &  \{-28.9145,-16.865,-10.9156,-3.11604+5.87462
 i,-3.11604-5.87462 i,-1.28288\} \\
\end{tabular}
\end{ruledtabular}
\end{table}

\subsection{Parameterization in terms of original fields}
\label{secz}

In this section, in contrast to the previous one, we keep the
potential parameterized in terms of the original fields $\phi_i$. 
This avoids the use of the chain rule together with tedious 
rewriting procedures and serves as a check of our results. As in the 
previous section, we expand the r.h.s.\ of Eq.\ (\ref{fleq}) and read
off flow equations for the couplings by comparison of coefficients, 
but now the expansion is in powers of the original fields $\phi_i$ 
instead of the invariants $\varphi_i,\gamma$. 
Again, in order to obtain the correct flow equations,  
accounting for all three anomaly terms, we have to keep at least 
three fields nonzero after having performed the second derivatives in 
Eq.\ (\ref{mm}). For a general rule which and how many fields one has
to keep at a minimum, in a case where the invariants are not known, 
we refer to Appendix \ref{o8}. \\
\noindent
For checking purposes we keep an additional field nonzero, say
$\pi_1$, and set $\pi_2$, $\pi_3$, $a_2$, and $a_3$ to zero after 
having computed the second derivatives. This means that
the comparison of coefficients is carried out using the potential
(\ref{ls}) for $\pi_2=\pi_3=a_2=a_3=0$ on the l.h.s.\ of the 
flow equation (\ref{fleq}). In this case the (scale-dependent) 
potential (\ref{ls}) reads
\begin{gather}
\label{uz}
U_k= a_1^2 m_{2,k}^2+\eta ^2 m_{2,k}^2+\sigma ^2 m_{1,k}^2
+\pi _1^2 m_{1,k}^2+\lambda _{\text{a$\eta $}} \left(a_1^4+\eta ^4\right) 
+\lambda _{\sigma \pi }  \left(\sigma ^4+\pi _1^4\right) 
\\ \nonumber
+\delta _1 \left(\pi _1^2 a_1^2+\eta ^2 \sigma ^2\right)
+\delta_2 a_1^2 \eta ^2+\delta_0  \left(a_1^2 \sigma ^2
+\pi _1^2 \eta ^2\right)+\kappa\pi _1 a_1 \eta  \sigma 
+\delta_3 \pi _1^2  \sigma ^2 \ ,
\end{gather}
\noindent
with
\begin{gather*}
 \lambda_{a\eta} \equiv \frac{\lambda_1}{24}-y+\frac{z}{2} \ , \ \ 
\lambda_{\sigma\pi} \equiv \frac{\lambda_1}{24}+y+\frac{z}{2} \ , \ \ 
\delta_0 \equiv \frac{\lambda_1}{12}+\lambda_2 -z  \ ,    \\  
\delta_1 \equiv \frac{\lambda_1}{12}-3z  \ , \ \   
\delta_2 \equiv \frac{\lambda_1}{12}+z-2y \ , \ \ 
\delta_3 \equiv \frac{\lambda_1}{12}+z+2y  \ , \ \ 
\kappa \equiv 4z+2\lambda_2 \ .
\end{gather*}
\noindent
Note that 
\begin{gather*}
 \delta_3 = 2 \lambda_{\sigma \pi} \ , \ \ 
\delta_2 = 2 \lambda_{a\eta} \ , \ \ 
\delta_0 = \delta_1 + \frac{\kappa}{2} \ , \ \ 
y = \frac{\lambda_{\sigma\pi}}{2}-\frac{\lambda_{a\eta}}{2} \ , \\
z=-\frac{\delta_1}{4}+\frac{\lambda_{\sigma\pi}}{4}
+\frac{\lambda_{a\eta}}{4} \ , \ \ 
\lambda_1=3\delta_1+ 9 \lambda_{a\eta} + 9 \lambda_{\sigma\pi} \ , \ \
\lambda_2 = \frac{\delta_1}{2} +\frac{\kappa}{2} 
-\frac{\lambda_{\sigma\pi}}{2}-\frac{\lambda_{a\eta}}{2} \ .
\end{gather*}
\noindent
We verified that we obtain unambiguous flow equations for 
$m_{1,k}^2$, $m_{2,k}^2$, $\lambda_{1,k}$, $\lambda_{2,k}$, $y_k$, 
and $z_k$, no matter from which of the coefficients in Eq.\ (\ref{uz}) 
we extract them (which is a freedom we have due to the additional
field we have kept nonzero). We do not state the flow equations and 
fixed points again; we have checked that they are equivalent to 
those found in Sec.\ \ref{sing}.

\subsection{Physical anomaly strength}
\label{pan}

So far we have considered only a finite anomaly strength. According to
Ref.\ \cite{Jungnickel:1995fp}, however, the limit 
$c \rightarrow -\infty$ should be closer to reality: 
in order to reproduce the correct 
vacuum mass of the eta meson in the two-flavor quark-meson model at 
tree-level, one has to choose a value for the anomaly strength 
($|c| \sim (958 \ {\rm MeV})^2$) which 
exceeds a physically reasonable UV cut-off scale for the
RG flow ($k \sim 600 \ {\rm MeV}$). Therefore, on all scales relevant
for the RG flow, effectively $c \rightarrow - \infty$.
More precisely, instead of this limit,
we should rather consider the limit 
\begin{gather} \label{goodlimit}
m_{2,k}^2= \frac{1}{2} \mu_k^2 -c_k \rightarrow \infty \ ,
\end{gather}
\noindent
otherwise $m_{1,k}^2 \equiv \frac{1}{2} \mu_k^2 +c_k \rightarrow
-\infty$ would impose severe constraints on the RG flow in order 
to finally obtain positive-definite masses for $\sigma$ and
$\vec{\pi}$.    \\
\noindent
In the limit $m_{2,k}^2 \rightarrow \infty$ the flow equations 
(\ref{flf})--(\ref{fll}) simplify to
\begin{gather}
k  \frac{\partial \bar{m}_1^2 }{\partial k}= 
-2 \bar{m}_1^2- \frac{4 \bar{l}_1}{\pi^2 \bar{\epsilon }_1^2} \ , \\
k  \frac{\partial \bar{l}_{12} }{\partial k} = 
-\bar{l}_{12} + \frac{8 \bar{l}_1 \left(\bar{l}_3 
+ 6 \bar{l}_{12} \right)}{3 \pi^2 \bar{\epsilon }_1^3} \ , \\
k  \frac{\partial \bar{l}_{1} }{\partial k} =  
-\bar{l}_1 + \frac{32 \bar{l}_1^2}{\pi^2 \bar{\epsilon }_1^3} \ , \\
k  \frac{\partial \bar{l}_{2} }{\partial k} =   
-\bar{l}_2 + \frac{2}{3 \pi^2} \frac{ \bar{l}_3^2 
+2 \bar{l}_3 \bar{l}_{12}+4\bar{l}_{12}^2}{\bar{\epsilon }_1^3} \ , \\
k  \frac{\partial \bar{l}_{3} }{\partial k} =  
-\bar{l}_3 +\frac{16 \bar{l}_3 \bar{l}_1 }{3 \pi^2 \bar{\epsilon }_1^3 }     \ .
\end{gather}
\noindent
The above flow equations have only one nontrivial fixed point, namely
the $O(4)$ fixed point,
\begin{gather}
\label{fpsol}
 \left(  \bar{m}_1^2 = -0.0555556,  \bar{l}_{1} = 0.216617,
   \bar{l}_{2} = 0 
, \bar{l}_{12} = 0 , \bar{l}_{3} = 0 \right) \ .
\end{gather}
\noindent
Calculating its stability-matrix eigenvalues,
\begin{gather}
\label{sev}
\{-1.77069,1.27069,-1,-0.83334,-0.5 \} \ ,
\end{gather}
\noindent
we find that it is IR unstable. According to standard rules one would,
erroneously, conclude that the phase transition cannot be of second
order. According to common sense, however, this cannot be true since 
the fields $\eta$ and $\vec{a}$ are infinitely heavy, so that 
fluctuations of these fields are completely suppressed and cannot 
affect the critical behavior. In Sec.\ \ref{coupvec} we explain, 
using a simpler model as an example, why we have to neglect the spurious 
negative eigenvalues when inferring the order of the phase
transition. From the discussion in Sec.\ \ref{coupvec}, we conclude 
that couplings occurring only in front of terms involving infinitely 
heavy fields have to be neglected in the stability analysis of fixed
points. This can be also understood from the fact that the
fluctuations represented by infinitely heavy fields are zero. \\
Inserting the fixed-point solution (\ref{fpsol}) 
into the rescaled potential (\ref{pinv}), we obtain 
\begin{gather}
 \bar{U}_{k=0} =  -0.0555556 \left( \bar{\sigma}^2 + 
\vec{\bar{\pi}}^2 \right) +  
0.216617 \left( \bar{\sigma}^2 + \vec{\bar{\pi}}^2 \right)^2 \ .
\end{gather}
\noindent
Since the fixed-point potential is $O(4)$ symmetric, we can choose 
$\vec{\bar{\pi}}_0 = 0$ in the vacuum state. Then, the rescaled vacuum
is given by
\begin{gather}
 (\bar{\sigma}_0 = 0.358099, \vec{\bar{\pi}}_0 = 0) \ .
\end{gather}
\noindent
Using these vacuum expectation values we calculate the rescaled mass 
eigenvalues (i.e., the rescaled physical masses):
\begin{gather}
\label{massinf}
 \bar{M}_{\sigma}^2 = 2/9 \ , \ \
 \bar{M}_{\pi_i}^2 = 0 \ , \ \
 \bar{M}_{\eta}^2 \rightarrow \infty \ , \ \
 \bar{M}_{a_i}^2 \rightarrow \infty \ .
\end{gather}
\noindent
We see that, as expected, we have three Goldstone bosons, 
the three pions $\vec{\pi}$, whereas $\eta$ and $\vec{a}$ are
infinitely heavy and thus decouple. 
Considering Eq.\ (\ref{pinv}), we conclude that the couplings 
$\bar{l}_{2}$, $\bar{l}_{12}$, and $\bar{l}_{3}$ appear only in 
front of terms involving infinitely heavy fields and must not be 
included in the stability analysis. Including only 
$\bar{m}_1^2$ and $\bar{l}_{1}$, we find the stability-matrix eigenvalues 
\begin{gather}
\label{sev1}
\{-1.77069,1.27069\}  \ ,
\end{gather}
\noindent
from which we finally conclude that there exists a 
stable $O(4)$ fixed point in case of infinite anomaly strength.
We also note that we verified that above the critical dimension, 
$d \geq 4$, the Gaussian fixed point becomes IR stable with mean-field
critical exponent $\nu = 1/2$, as expected.

\section{Coupled vector model}
\label{coupvec}

In order to justify why we can neglect the spurious negative
eigenvalues occurring in Sec.\ \ref{pan}, we discuss a simpler 
model where the reasons become transparent. We consider the 
case of the most simple coupled vector model, which involves two scalar fields 
$\phi_1$ and $\phi_2$:
\begin{gather}
U = m_1^2 \phi_1^2+m_2^2 \phi_2^2+\frac{\lambda_{11}}{24} \phi_1^4 
+ \frac{\lambda_{12}}{12} \phi_1^2 \phi_2^2 
+ \frac{\lambda_{22}}{24} \phi_2^4 \ .
\end{gather}
\noindent
For a mean-field analysis of the model we refer to Ref.\
\cite{Galam:1988}, 
for a leading-order $\epsilon$-expansion to Ref.\ \cite{Aharony:1976br}. \\
\noindent
Using the method of Taylor expansion and comparison of
coefficients, we find the following flow equations:
\noindent
\begin{gather}
k  \frac{\partial \bar{m}_1^2 }{\partial k}= 
-2 \bar{m}_1^2-\frac{1}{36 \pi ^2}\left(
\frac{3 \bar{\lambda }_{11}}{\bar{\epsilon }_1^2}
+\frac{\bar{\lambda }_{12}}{\bar{\epsilon }_2^2}\right) \ , \ \
k  \frac{\partial \bar{m}_2^2 }{\partial k}=  
-2 \bar{m}_2^2-\frac{1}{36 \pi ^2}\left(
\frac{3 \bar{\lambda }_{22}}{\bar{\epsilon }_2^2}
+\frac{\bar{\lambda }_{12}}{\bar{\epsilon }_1^2}\right) \ , \\
k  \frac{\partial \bar{\lambda}_{11} }{\partial k}=
   -\bar{\lambda }_{11}+\frac{1}{\pi ^2}\left(
\frac{ \bar{\lambda }_{11}^2}{\bar{\epsilon }_1^3}
+\frac{ \bar{\lambda }_{12}^2}{9 \bar{\epsilon}_2^3}\right) \ , \ \
k  \frac{\partial \bar{\lambda}_{22} }{\partial k}=
-\bar{\lambda }_{22}+\frac{1}{\pi ^2}\left(
\frac{ \bar{\lambda }_{22}^2}{\bar{\epsilon }_2^3}
+\frac{ \bar{\lambda }_{12}^2}{9 \bar{\epsilon}_1^3}\right) \ , \\
k  \frac{\partial \bar{\lambda}_{12} }{\partial k}=
-\bar{\lambda }_{12}  + \frac{\bar{\lambda }_{12}}{9 \pi ^2 
\bar{\epsilon }_1^3 \bar{\epsilon }_2^3} 
\left[2 \bar{\lambda }_{12} \left(\bar{\epsilon}_1+\bar{\epsilon}_2\right) 
\bar{\epsilon }_2 \bar{\epsilon }_1+3 \bar{\lambda }_{22} 
\bar{\epsilon }_1^3+3 \bar{\lambda }_{11} \bar{\epsilon }_2^3\right]  \ , 
\end{gather}
\noindent
where we again used the abbreviation (\ref{abbr}). \\
\noindent
In this work we are only interested in the Ising fixed point,
\begin{gather}
 ( \bar{m}_1^2 = -0.03846 ,  \bar{m}_2^2 = 0  ,  
\bar{\lambda}_{11} = 7.76271  ,  \bar{\lambda}_{12}=0  ,  
\bar{\lambda }_{22}=0) \ ,
\end{gather}
\noindent
the stability-matrix eigenvalues of which,
\begin{gather}
 \{  -2  ,  -1.84256  ,  1.1759  ,  -1  ,   -0.666667 \} \ ,
\end{gather}
\noindent
indicate that it appears to be unstable in the $ \bar{m}_2^2$, $\bar{\lambda}_{12}$,
and $\bar{\lambda }_{22}$ directions. \\
\noindent
Examining the above flow equations in the limit $\bar{m}_2^2 \rightarrow \infty$,
\begin{gather}
k  \frac{\partial \bar{m}_1^2 }{\partial k}= 
-2 \bar{m}_1^2-\frac{1}{12 \pi ^2} \frac{ 
\bar{\lambda }_{11}}{\bar{\epsilon }_1^2} \ , \ \
 k  \frac{\partial \bar{\lambda}_{11} }{\partial k}= 
  -\bar{\lambda }_{11}+\frac{1}{\pi ^2} \frac{ 
\bar{\lambda }_{11}^2}{\bar{\epsilon }_1^3} \ , \\
k  \frac{\partial \bar{\lambda}_{22} }{\partial k}=
-\bar{\lambda }_{22}+\frac{1}{9 \pi ^2}\frac{ 
\bar{\lambda }_{12}^2}{ \bar{\epsilon}_1^3} \ , \ \
 k  \frac{\partial \bar{\lambda}_{12} }{\partial k}=
-\bar{\lambda }_{12}  +  \frac{1}{3 \pi ^2 } \frac{
\bar{\lambda }_{12} \bar{\lambda }_{11}}{\bar{\epsilon }_1^3 }    \ , 
\end{gather}
\noindent
we still find negative eigenvalues corresponding to the unstable 
$\bar{\lambda}_{12}$ and $\bar{\lambda }_{22}$ directions, respectively. 
Obviously, we have the same situation as in Sec.\ \ref{pan}. Formally, 
the negative eigenvalues would indicate that the Ising fixed point 
is IR unstable. In this particular case, however, one cannot conclude 
from this that the phase transition is fluctuation-induced first
order. Fluctuations in $\bar{\lambda}_{12}$ and $\bar{\lambda }_{22}$
direction are completely suppressed due to the infinitely heavy 
$\phi_2$ field and cannot affect the critical behavior. 
To prove this, we investigate in detail the scale evolution of the 
dimensionful potential for different initial values for the 
parameters in the UV. Using the invariants
\begin{gather}
 \varphi_1 = \phi_1^2 \ , \ \ \varphi_2 = \phi_2^2 \ ,
\end{gather}
\noindent
we make the following ansatz for the potential running under the RG flow:
\begin{gather}
 U_k = V_k(\varphi_1) + W_k(\varphi_1) \varphi_2 + X_k(\varphi_1) \varphi_2^2 \ .
\end{gather}
\noindent
Having expressed the mass eigenvalues $M_i^2$ in terms of 
$\varphi_1$ and $\varphi_2$, we expand the r.h.s.\ of Eq.\
(\ref{fleq}) 
and read off flow equations for $V_k(\varphi_1)$, $W_k(\varphi_1)$, 
and $X_k(\varphi_1)$ by comparison of coefficients. We solve the 
resulting system of three partial differential equations together 
with the initial conditions
\begin{gather}
 V_{k= \Lambda} (\varphi_1) = m_{1,\Lambda}^2 \varphi_1
+\frac{\lambda_{11,\Lambda}}{24} \varphi_1^2 \ , \ \ 
W_{k=\Lambda} (\varphi_1) =m_{2,\Lambda}^2 
+\frac{\lambda_{12,\Lambda}}{12} \varphi_1 \ , \ \ 
X_{k=\Lambda}(\varphi_1) = \frac{\lambda_{22,\Lambda}}{24} \ .
\end{gather}
\begin{figure}[htbp]
    \includegraphics[scale=0.42]{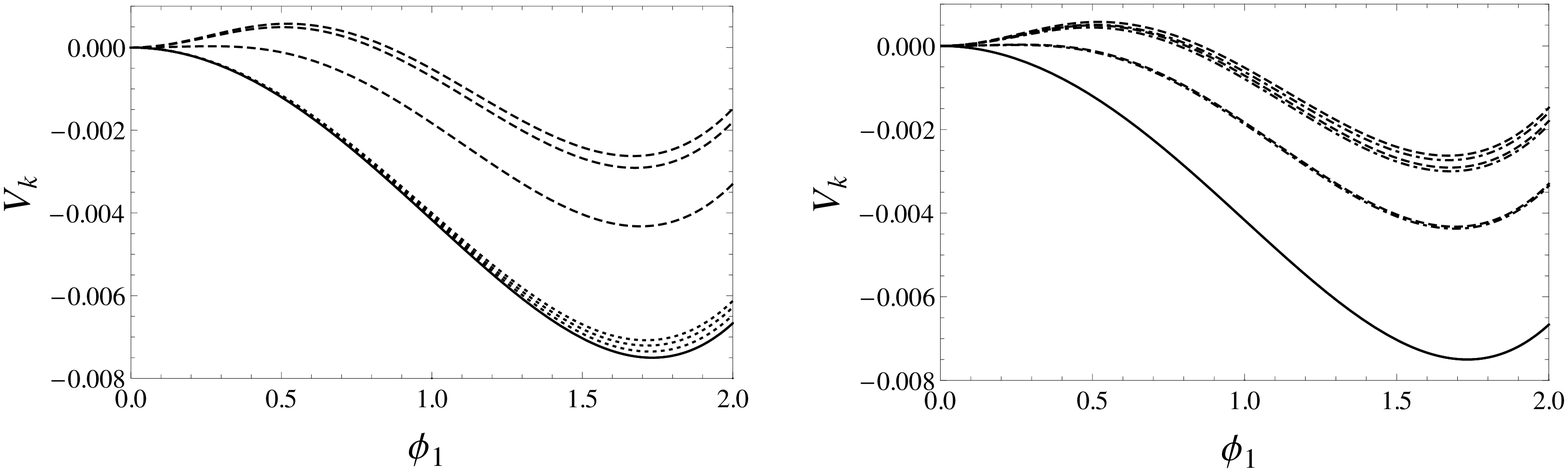}
    \includegraphics[scale=0.42]{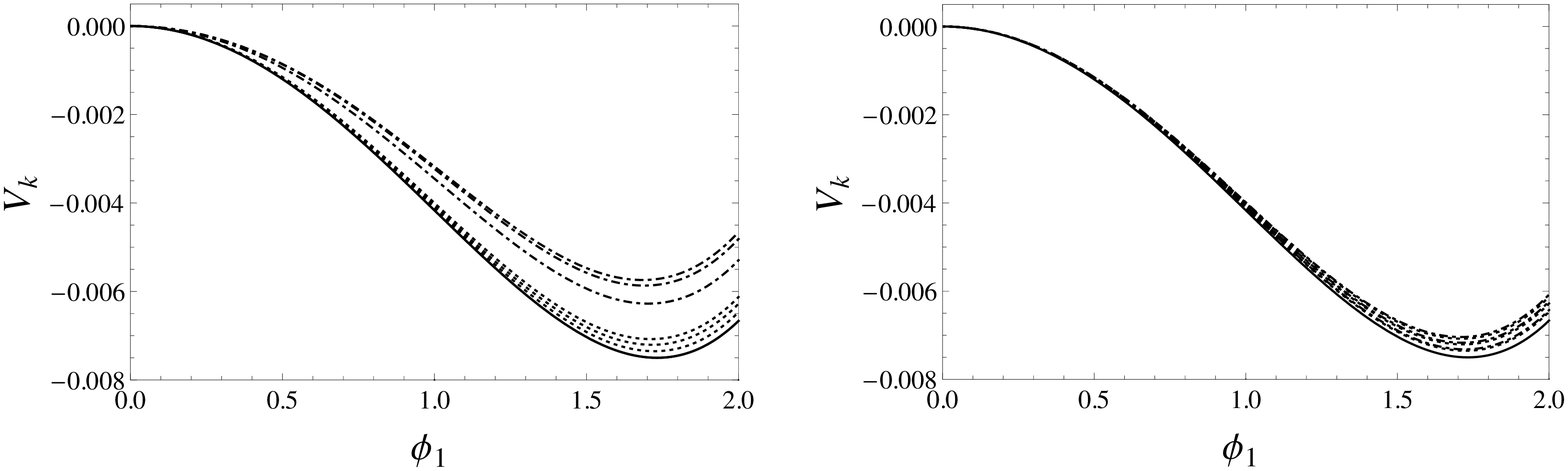} 
  \caption{Scale evolution of the potential $V_k$. In each panel,
the solid line is the same and corresponds to the start of the evolution
in the UV ($k/\Lambda = 1$). Furthermore, in each panel 
there are two sets of three curves (drawn with identical line mode). 
These three curves correspond to the RG
potentials at the scales $k/\Lambda=0.7$, $k/\Lambda=0.4$, and
$k/\Lambda=0.12$ respectively. For all panels, 
$m_{1,\Lambda}^2= -0.005 \Lambda^2$, $\lambda_{11,\Lambda} = 0.02
\Lambda$, $\lambda_{22,\Lambda}=0$. In the upper left, the lower left,
and the lower right panel, the dotted curves correspond to 
$m_{2,\Lambda}^2=0$, $\lambda_{12,\Lambda}=0$ (and therefore coincide
with solutions for the Ising model).
In the upper left and upper right panel,
the dashed curves are for $m_{2,\Lambda}^2=0$, $\lambda_{12,\Lambda}=
8 \Lambda$. In the lower left 
panel, the dot-dashed curves are
for $m_{2,\Lambda}^2=0.5 \Lambda^2$ and $\lambda_{12,\Lambda}=8
\Lambda$. In the lower right panel, 
the dot-dashed curves are
for $m_{2,\Lambda}^2=8 \Lambda^2$ and $\lambda_{12,\Lambda}=8
\Lambda$.
 }
  \label{F1}
\end{figure}

\noindent
Figure \ref{F1} illustrates the potential for
various values of the RG flow parameter $k$ for various values
of $m_{2, \Lambda}^2$ and $\lambda_{12,\Lambda}$ for
fixed values of $m_{1,\Lambda}^2$, $\lambda_{11,\Lambda}$, and
$\lambda_{22,\Lambda}=0$. We observe that
the influence of the coupling $\bar{\lambda}_{12}$ 
on the shape of the potential becomes smaller for larger values of $m_2^2$. 
We have checked that the same is true for nonzero
values of the coupling $\bar{\lambda}_{22}$. We also observe
that the RG-evolved potential exhibits the typical shape 
for a (fluctuation-induced) first-order phase transition
in the case of a light $\varphi_2$ field (upper panels), while
the transition remains of second order for a heavy $\varphi_2$ field
(lower panels).

\section{Conclusions}
\label{conclusions}

We investigated the conjecture that the two-flavor chiral phase
transition of QCD can be of second order in the presence of the 
axial anomaly. We studied the most general renormalizable Lagrangian 
invariant under $SU(2)_A \times U(2)_V$, using the FRG method 
in the local-potential approximation. We took into account all 
possible 't Hooft determinant-like terms, the couplings of which 
we denoted as $c$, $y$, and $z$, respectively. We distinguished 
between the case of finite and the limit of divergent anomaly strength
$c$.  \\
\noindent
Our conclusions are as follows. An $O(4)$ IR fixed point indeed
exists for the two-flavor linear sigma model in the presence of the 
axial anomaly. However, it is only IR stable in the case of infinite 
anomaly strength. This case is reasonable if the IR value of the 
anomaly strength exceeds the cut-off scale of the linear sigma model, 
which is true at the mean-field level but not beyond this admittedly very
crude approximation. For finite anomaly strength, however, we found 
that the $O(4)$ IR fixed point is unstable. 
Nevertheless, we find other IR stable fixed points which are in
the $SU(2) \times U(2)$ universality class. These have
unphysical mass-matrix eigenvalues in our fourth-order truncation of
the potential and were thus neglected in our considerations. However,
in a scheme which accounts for higher orders, they might become
physical \cite{Marainprep}, indicating that 
the two-flavor chiral phase transition of QCD could be of second order,
but not with $O(4)$ critical exponents. We want to note that the possibility of another
universality class ($U(2)_L \times U(2)_R / U(2)_V$) has also been recently emphasized
in Ref.\ \cite{PhysRevD.86.114512}. 
On the other hand, if the $SU(2) \times U(2)$ fixed points should remain unphysical,
the absence of other IR stable fixed points indicates that
the phase transition should be fluctuation-induced first order. 
However, we note that the 
strength of the first-order phase transition depends on the 
initial values for the parameters in the UV and could be
extremely weak, which would make it practically indistinguishable
from a second-order phase transition in this case \cite{Marainprep}.

\section*{Acknowledgment}

M.G.\ would like to thank HGS-HIRe for FAIR for funding. The authors would like
to thank Bengt Friman,  Francesco Giacosa, Holger Gies, 
Bertram Klein, Mario Mitter,
Jan M.\ Pawlowski, Rob Pisarski, Bernd-Jochen Schaefer,  
and Stefan Schramm for valuable discussions.

\appendix

\section{Constructing invariants} \label{AppA}

In condensed-matter
systems, finite groups $G$ play an important role. For such groups,
there exist practical methods how to construct the most 
general $G$-invariant
Landau-Wilson polynomials for certain representations
$\Gamma(G)$. These methods have been applied to study
phase transitions in various condensed-matter systems 
\cite{Toledano:1985,Kim:1986,Hatch:1986,Stokes:1987}.
For arbitrary continuous groups, however, such a program is, 
at the very least, not well documented. In the following 
we describe how to construct
the $SU(2)_A \times SU(2)_V$ invariants for the 
$[\bar{2},2]+[2,\bar{2}]$ representation. We note that our
method is not restricted to this special case, and we have
checked that it can be successfully applied to other groups as well. 
However, one has to know the explicit form of the symmetry 
transformation for the representation
of interest. \\
\noindent
The $[\bar{2},2]+[2,\bar{2}]$ representation is $8$-dimensional. 
Accordingly, the corresponding invariants of order $N$ are 
polynomials in eight components which are in our notation the 
fields $\sigma$, $\vec{\pi}$, $\eta$, and $\vec{a}$, i.e.,
they are of the form
\begin{gather}
 p = \sum_{m_i \in m} c_i m_i \ , \label{poly}
\end{gather}
\noindent
where $m$ denotes the set of all possible monomials of order $N$,
\begin{gather}
 m = \{ \sigma^{n_1} \pi_1^{n_2} \pi_2^{n_3} \pi_3^{n_4} 
\eta^{n_5} a_1^{n_6} a_2^{n_7}  a_3^{n_8}  \} \ , 
\ \ n_i \in \mathbb{N} \ , \ \ \sum_i n_i = N \ ,
\end{gather}
\noindent
and the coefficients $c_i$ are expected to be rational multiples of 
each other. \\
\noindent
Infinitesimal $SU(2)_A$ transformations for the above 
representation are determined by \cite{Toki}
\begin{gather}
 \sigma' = \sigma + \vec{\alpha} \cdot \vec{\pi} \ , \ \ 
\pi_i' = \pi_i -\alpha_i \sigma \ , \ \ 
 \eta' = \eta - \vec{\alpha} \cdot \vec{a} \ , \ \ 
a_i' = a_i +\alpha_i \eta \ , \label{inftr}
\end{gather}
\noindent
where $\vec{\alpha}=(\alpha_1,\alpha_2,\alpha_3)$ consists of 
three infinitesimal angles. 
Infinitesimal $SU(2)_V$ transformations for the above 
representation are determined by
\begin{gather}
 \sigma' = \sigma  \ , \ \ 
\vec{\pi}' = \vec{\pi} + \vec{\beta} \times \vec{\pi}  \ , \ \ 
 \eta' = \eta  \ , \ \ 
\vec{a}' = \vec{a} + \vec{\beta} \times \vec{a} \ , \label{inftru2v}
\end{gather}
\noindent
where $\vec{\beta}=(\beta_1,\beta_2,\beta_3)$ consists of 
three infinitesimal angles. \\
\noindent
Under the transformation (\ref{inftr}), the polynomial $p$ transforms as 
\begin{gather}
 p \rightarrow p'  = \sum_{m_i \in m} c_i'(\vec{c},\vec{\alpha}) m_i \ ,
\end{gather}
\noindent
where the new coefficients, $c_i'$, depend on the coefficients 
$\vec{c}$ and the angles $\vec{\alpha}$, 
and where we only keep terms linear in $\alpha_i$.
Since invariants are defined by $p=p'$, we obtain a system of equations,
\begin{gather}
 c_i = c_i'(\vec{c},\vec{\alpha}) \ , \label{sysin}
\end{gather}
\noindent
determining all invariants of order $N$. \\
\noindent
For $N=2$, the sum in Eq. (\ref{poly}) runs from $i=1$ to $i=36$, 
since there are $36$ different monomials of order $N=2$.
Using for example \textit{Mathematica's} option \textit{SolveAlways} 
\cite{Mathematica}, solutions for
the coefficients $c_i$ can be found, such that Eqs.\ (\ref{sysin}) 
are fulfilled for arbitrary
values of the angles $\alpha_i$. Inserting the solution into the 
general ansatz (\ref{poly}), we obtain
\begin{gather}
 p = c_1 ( \sigma^2 + \vec{\pi}^2) + c_2 ( \eta^2  + \vec{a}^2) 
+ c_3 ( \sigma \eta - \vec{\pi} \cdot \vec{a}) \ .
\end{gather}
\noindent
Since the coefficients $c_i$ are independent from each other, 
there exist exactly three linearly independent invariants of order
$N=2$:
\begin{gather}
\label{In2}
 \varphi_1 =  \sigma^2 + \vec{\pi}^2 \ , \ \  
\varphi_2 = \eta^2  + \vec{a}^2 \ , \ \ 
\varphi_3 = \sigma \eta - \vec{\pi} \cdot \vec{a} \ .
\end{gather}
\noindent
For $N=4$, the sum in Eq.\ (\ref{poly}) runs from $i=1$ to $i=330$, 
since there are $330$ different monomials of order $N=4$.
Again, using \textit{Mathematica}, we find solutions for
the coefficients $c_i$, such that Eqs.\ (\ref{sysin}) are fulfilled 
for arbitrary
values of the angles $\alpha_i$. Inserting the solution into 
the general ansatz (\ref{poly}), we obtain 
\begin{gather}
 p = c_1 \left(\eta ^2+ \vec{a}^2\right)^2   
+c_2 \left(\sigma ^2+ \vec{\pi}^2\right)^2   
+ c_3 \left(-  \sigma \eta +\vec{\pi} \cdot \vec{a}\right)^2 
+c_4 \left(-\sigma \eta +\vec{\pi} \cdot \vec{a}\right) 
\left(\sigma ^2+ \vec{\pi}^2\right) \nonumber\\
 + c_5 \left(\eta ^2+ \vec{a}^2\right) 
\left(-\sigma \eta +\vec{\pi} \cdot \vec{a}\right)
+c_6 \left[ \left(\eta ^2+ \vec{a}^2\right) 
\left(\sigma ^2+ \vec{\pi}^2\right)
-\left(\sigma \eta -\vec{\pi} \cdot \vec{a}\right)^2 \right] \ .
\end{gather}
\noindent
Since the coefficients $c_i$ are independent from each other, 
there exist exactly four linearly independent invariants of order
$N=4$:
\begin{gather}
\label{In4}
 \varphi_1^2 \ , \ \  \varphi_2^2  \ , \ \ \varphi_1 \varphi_2 \ , \ \
 \gamma = \varphi_3^2 \ .
\end{gather}
\noindent
Note that the quadratic invariant $\varphi_3$ is not invariant under 
parity transformations
\begin{gather}
 \sigma \rightarrow \sigma \ , \ \ \vec{\pi} \rightarrow \vec{\pi} \ , \ \ \eta \rightarrow -\eta \ , \ \ \vec{a} \rightarrow -\vec{a} \ ,
\end{gather}
\noindent
and therefore cannot appear in a theory without parity violation. \\
\noindent
Note further that the invariants (\ref{In2}) and (\ref{In4}) are also invariant under
$SU(2)_V$ transformations (\ref{inftru2v}). 
Proceeding along the same lines described above one can derive
several additional invariants for this symmetry. Since these
are not $SU(2)_A$ symmetric, and hence no
$SU(2)_A \times SU(2)_V$ invariants, we do not list them here.

\section{Critical Exponents from the Stability Matrix}
\label{crex}

In the following we describe how to calculate critical exponents 
proceeding in complete analogy to Ref.\ \cite{Litim:2002cf}. 
The method is appropriate as long as the anomalous dimension $\eta$ 
is small, which is assumed to be the case in the local-potential 
approximation. For given beta functions $\beta_i(\bar{p}) 
\equiv k \partial_k \bar{p}_i$ for the rescaled parameters 
$\bar{p} = \{\bar{p}_i \}$ (i.e., the rescaled mass terms and
couplings) of the Lagrangian, the stability matrix for a fixed 
point is defined as
\begin{gather}
 (S_{ij}) \equiv \left( \frac{\partial \beta_i}{\partial \bar{p}_j} 
\right) \Bigl\vert_{\bar{p}=\bar{p}*} \ ,
\end{gather}
\noindent
where a fixed point $\{ \bar{p}_i^* \}$ is determined by
\begin{gather}
 \beta_i (\{ \bar{p}_i^* \}) =0 \ .
\end{gather}
The stability properties of a fixed point can be determined from the 
eigenvalues of the stability matrix $S$. Eigenvalues with positive 
real part correspond to IR stable (UV unstable) 
directions, whereas eigenvalues with negative real part correspond 
to IR unstable (UV stable) directions. If a fixed point is IR (UV) 
stable in a direction in coupling space, the 
flow, for decreasing $k$, in the neighbourhood of the fixed point 
is directed towards it (away from it) in this direction in coupling
space. For fixed points associated with second-order phase
transitions, for every plane in coupling space an IR stable 
direction exists. However, since a phase transition always 
requires that a scaling variable (e.g., the temperature $T$)
approaches a critical value, there has to exist at least one IR 
unstable direction. Tuning a system towards the critical point 
corresponds to tuning the parameters $\bar{p}_i$ to a point 
on the critical surface (a point which is attracted by the 
IR fixed point). Such a fixed point can be associated with a 
second-order phase transition and is simply called IR stable. 
In case of a single scaling variable, the eigenvalues of
the stability matrix for an IR stable fixed point have positive 
real parts, except for one which is negative, say $y_1$. 
The critical exponent $\nu$, determined by
\begin{gather}
 T \rightarrow T_c \ : \ \ \xi \sim |T-T_c|^{-\nu} \ ,
\end{gather}
is then given by
\begin{gather}
\label{nue}
 \nu = -\frac{1}{y_1} \ .
\end{gather}

\section{$O(N)$ IR Fixed Points}
\label{o8}

\begin{table}
\caption{\label{tab1}Stability-matrix eigenvalues, $y_i$, 
for the Wilson-Fisher fixed point of the $O(N)$ model, $d=3$, 
FRG (in local-potential approximation), 
up to quartic coupling. The bar denotes rescaled quantities.}
\begin{ruledtabular}
\begin{tabular}{l|l l l l l l l l}
$N$ & $\frac{1}{2} \bar{\mu}_*^2$ & $\bar{\lambda}_{1*}$ 
& $\nu = -1/y_1$ & $y_2$  \\ \hline
1 & -0.03846 & 7.76271 & 0.54272=-1/-1.84256 & 1.1759  \\
2 & -0.04545 & 6.67366 & 0.55149=-1/-1.81327 & 1.21327  \\
4 & -0.05556 & 5.1988 & 0.564751=-1/-1.77069 & 1.27069  \\
8 & -0.06757 & 3.59143 & 0.581495=-1/-1.71971 & 1.34471  \\
\end{tabular}
\end{ruledtabular}
\end{table}

In order to determine the universality class that a 
fixed point belongs to, one has to compare its stability-matrix 
eigenvalues with those of the fixed points defining certain 
universality classes. In Table \ref{tab1} we list stability-matrix 
eigenvalues for $O(N)$ models,
\begin{gather}
 U = \frac{1}{2} \mu^2 \sum_{n=1}^{N} \phi_i^2 
+ \frac{\lambda_1}{24} \left(\sum_{n=1}^{N} \phi_i^2 \right)^2 \ ,
\end{gather}
\noindent
which are relevant for our discussion. We keep only terms with 
relevant canonical scaling dimension. Higher-order terms are 
irrelevant with respect to the Gaussian fixed point and should 
be negligible also with respect to nontrivial fixed points in a 
resummed $\epsilon$-expansion. This is different in the FRG approach. 
At non-trivial fixed points, higher-order terms are expected to have 
a distinct effect on critical exponents \cite{Litim:2002cf}. 
However, in this work we are only interested in identifying 
the universality class of a fixed point, so we can neglect these 
higher orders, if we also do this in the model which
we want to compare the $O(N)$ model with. For more evolved
FRG studies of $O(N)$ models we refer for example to
Refs.\ \cite{Berges:1997eu,BJ2000,Litim:2002cf,Braun:2008sg,Friman2010,Kamikado:2013sia}
and references therein. \\
\noindent
As an example, we can now explicitly determine the 
universality class of the IR fixed point of the flow equations 
(69)--(71) in Ref.\ \cite{Fukushima:2010ji}. These are the flow 
equations we obtain from our results in the limit $c,y,z \rightarrow
0$. The IR fixed point is given by 
$(\bar{\mu}_*^2,\bar{\lambda}_{1*},\bar{\lambda}_{2*})=(-0.135,3.591,0)$
and is unstable as one infers from the stability-matrix 
eigenvalues $\{-1.71971,1.34471,-0.25 \}$, which are in perfect 
agreement with the $O(8)$-values in Table \ref{tab1}. \\
\noindent
In the context of $O(N)$ models we can easily study the 
influence of setting fields to zero after having performed 
the second derivatives in Eq.\ (\ref{mm}). It does not affect the 
results at all, because the coefficients in the expansion of the 
flow equation (\ref{fleq}) in terms of fields do not change by 
setting certain fields to zero. Due to the $O(N)$ symmetry we can 
read off the flow equation for $\lambda_1$ from any quartic 
term. This argument directly generalizes to all other potentials 
for which one obtains unambiguous flow equations for all couplings, 
keeping all fields nonzero. In such a case, instead of keeping all 
fields nonzero, one can set as many fields to zero 
[after having performed the second derivatives in Eq.\ (\ref{mm})] 
as one likes, as long as one still obtains a flow equation 
for each coupling. The flow equations are the same in both cases.

\bibliographystyle{unsrt}
\bibliography{mybib_RG}


\end{document}